\documentclass[
superscriptaddress,
nofootinbib,
 amsmath,amssymb,
 aps,
pra,
notitlepage,
longbibliography
]{revtex4-1}

\usepackage{dcolumn}
\usepackage{bm}
\usepackage{epsfig}
\usepackage{graphicx}
\usepackage{latexsym}
\usepackage{amsfonts}
\usepackage{float} 
\usepackage{setspace}
\usepackage{soul}
\usepackage{graphicx}
\usepackage{amsmath}
\usepackage{mathrsfs}
\usepackage{verbatim}
\usepackage{color}
\usepackage{SIunits}
\usepackage{hyperref}
\usepackage{physics}
\usepackage{tabularx}
\usepackage{array}

\newcommand{\be}{\begin{equation}}
\newcommand{\ee}{\end{equation}}
\newcommand{\ba}{\begin{array}}
\newcommand{\ea}{\end{array}}
\newcommand{\bqa}{\begin{eqnarray}}
\newcommand{\eqa}{\end{eqnarray}}

\begin{document}
\title{Fidelity and efficiency analysis for heralded entanglement swapping in lossy channels: linear and nonlinear optical approaches}

\author{Kejie Fang} 
\email{kfang3@illinois.edu}
\affiliation{Holonyak Micro and Nanotechnology Laboratory and Department of Electrical and Computer Engineering, University of Illinois at Urbana-Champaign, Urbana, IL 61801 USA}
\affiliation{Illinois Quantum Information Science and Technology Center, University of Illinois at Urbana-Champaign, Urbana, IL 61801 USA}
\author{Elizabeth A. Goldschmidt} 
\affiliation{Illinois Quantum Information Science and Technology Center, University of Illinois at Urbana-Champaign, Urbana, IL 61801 USA}
\affiliation{Department of Physics, University of Illinois at Urbana-Champaign, Urbana, IL 61801 USA}

\begin{abstract} 
Bell state measurements (BSMs) of photonic qubits are used for quantum networking protocols to herald the distribution and transfer of quantum information. However, standard BSMs based on linear optics (LO-BSMs) require identical photons and are susceptible to errors from multiphoton emissions, leading to reduced fidelity of protocols, particularly in the absence of postselection. To overcome these challenges, Bell state measurements based on nonlinear optics (NLO-BSMs) have been proposed and implemented, leveraging sum-frequency generation (SFG) to filter out multiphoton emissions and improve the fidelity without postselection. Here we analyze the fidelity of LO-BSM and NLO-BSM heralded entanglement swapping in lossy channels and compare their performance under realistic conditions. We also explore the impact of state-of-the-art nonlinear photonics platforms on SFG efficiency and highlight how advances in nanophotonics will enable practical, high-fidelity NLO-BSM-based quantum networking applications.
\end{abstract} 

\maketitle

\section{Introduction}

Many quantum networking protocols, such as quantum teleportation and entanglement swapping, depend on linear-optical Bell state measurements (LO-BSMs), which use interference between two identical photons at a beamsplitter to facilitate the heralding of quantum information transfer and distribution. However, LO-BSMs require the two input photons to be identical, which can lead to false heralds in the presence of multiphoton emission \cite{kok2000postselected,pan2003experimental}---a common occurrence in probabilistic quantum light sources and many quantum emitters. For entanglement swapping, postselection on exactly one detected photon at each end of the final, swapped pair (i.e., undetected outer photons in Fig. \ref{fig:eswapping}a) is used to overcome multiphoton-based infidelity, but it precludes any subsequent manipulation or use of the photons and restricts the usefulness of any quantum protocol. Without postselection, LO-BSMs and standard photon pair sources have fundamentally restricted fidelity for any entanglement distribution-based protocol \cite{sangouard2011quantum,azuma2023quantum}.  The requirement for photon indistinguishability imposes another fundamental constraint on LO-BSMs, as any deviation from perfect indistinguishability reduces the protocol's fidelity. In practical quantum networks, where single and entangled photons are independently generated and transmitted through dispersive or unstable channels, ensuring perfect indistinguishability is a significant challenge. Consequently, real-world demonstrations of quantum information transfer or entanglement distribution over even a small number of nodes suffer from substantial errors  \cite{sun2016quantum,sun2017entanglement,shen2023hertz,knaut2024entanglement}. Additionally, a LO-BSM can only distinguish two of the four Bell states without a dramatic increase in the system complexity via the addition of ancillary photons, and even with ancillary photons, it cannot distinguish all Bell states \cite{lutkenhaus1999bell}.

To address these limitations, nonlinear-optical Bell state measurements (NLO-BSMs) have been proposed as a solution for non-postselected quantum teleportation \cite{kim2001quantum}  and entanglement swapping \cite{sangouard2011faithful}. Unlike LO-BSMs, NLO-BSMs utilize sum-frequency generation (SFG) between non-degenerate photons to filter out multiphoton emissions, thereby enabling quantum teleportation and entanglement swapping with a high non-postselected fidelity. This enables quantum light sources to be driven more efficiently with less reduction in the fidelity of protocols. The NLO-BSM also avoids the stringent requirement of identical photons, as the two input photons now only need to satisfy the phase-matching condition of the SFG process. Deviating from this condition affects only the efficiency but not the fidelity of the protocols, as the phase-matching condition filters the variations of input photons. However, the efficiency of NLO-BSMs is fundamentally constrained by the typically small efficiency of SFG processes. Bulk nonlinear crystals and waveguides, which have been the primary platform for SFG to date, typically require intense optical fields for SFG and other nonlinear processes  \cite{kim2001quantum,sephton2023quantum,qiu2023remote,tanzilli2005photonic,sangouard2011faithful,fisher2021single,guerreiro2013interaction,guerreiro2014nonlinear},  although we note the recent work demonstrating entanglement swapping at the single-photon level in such a platform \cite{tsujimoto2024experimental}. Integrated photonics, leveraging low optical losses and much tighter light confinement in nanophotonic structures, present a pathway toward much larger nonlinearity-to-loss ratios for single-photon operations \cite{lu2020toward,zhao2022ingap}. As a result, SFG probabilities on the order of $4 \times 10^{-5}$ have been demonstrated using nanophotonic cavities, enabling NLO-BSM heralded quantum teleportation at the single-photon level \cite{akin2025faithful}. We show below how non-postselected NLO-BSM-based entanglement swapping can be performed at high fidelity, unlike LO-BSM-based entanglement swapping which is limited to a fidelity of 1/3 for non-postselected operation. We note that LO-BSM can be used in combination with heralded entanglement generation \cite{sliwa2003conditional,pittman2003heralded,barz2010heralded,wagenknecht2010experimental} to perform high fidelity (at least in theory) non-postselected entanglement swapping. But this comes with a substantial reduction in rate due to the need to generate two ancillary photon pairs for every entangled pair needed, i.e., total six photons. Thus, even at $4 \times 10^{-5}$ efficiency, a NLO-BSM-based scheme can operate at a much higher rate than LO schemes based on heralded entanglement sources  \cite{sangouard2011faithful}. 

In this paper, we analyze the non-postselected fidelity of heralded entanglement swapping based on LO-BSMs and NLO-BSMs, respectively, in the presence of lossy channels. We derive general results for both approaches under arbitrary loss conditions and compare them for practical scenarios. Additionally, we examine the efficiency of SFG in  state-of-the-art nonlinear photonics platforms. We argue that the high fidelity achievable with NLO-BSMs, along with improved SFG efficiency enabled by advances in nonlinear nanophotonics, positions NLO-BSM heralded entanglement swapping as a promising approach for practical quantum applications.

\section{Linear-optical Bell state measurement}

\begin{figure}[!htb]
\begin{center}
\includegraphics[width=1\columnwidth]{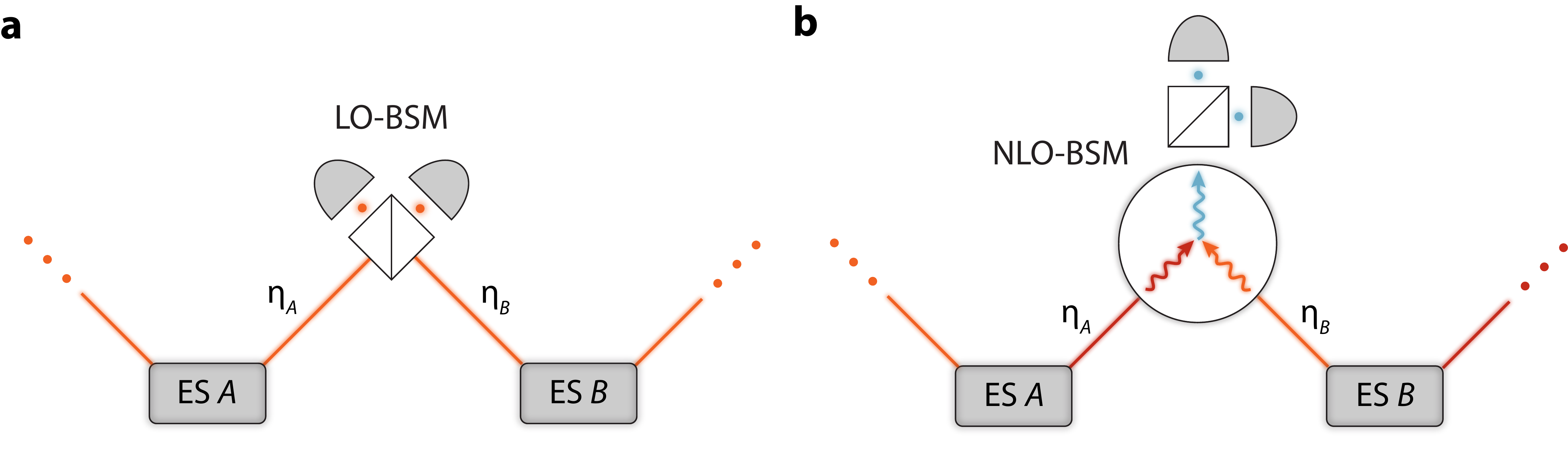}
\caption{\textbf{Entanglement swapping.} \textbf{a}. Entanglement swapping via linear-optical Bell state measurement (LO-BSM). \textbf{b}. Entanglement swapping via nonlinear-optical Bell state measurement (NLO-BSM). ES: entanglement source. $\eta_A$ and $\eta_B$ are the loss of the two channels. }
\label{fig:eswapping}
\end{center}
\end{figure}

We consider entanglement swapping between two spontaneous parametric down-conversion (SPDC) sources $A$ and $B$ (Fig. \ref{fig:eswapping}). The state of the the down-converted photons of each source can be written as \cite{braunstein2005quantum}:
\begin{equation}
	\begin{gathered}
		\vert \psi_\mathrm{SPDC} \rangle = \sqrt{1-\epsilon}\sum_{n = 0}^{\infty} \epsilon^{n/2} \vert n \rangle_s\vert n \rangle_i.
	\end{gathered}
	\label{eq:SPDC}
\end{equation}
where $\epsilon$ represents the pump conversion efficiency and we have chosen $\epsilon$ to be real without loss of generality.  The single-photon-pair probability is given by $p=(1-\epsilon)\epsilon\leq\frac{1}{4}$ and we can rewrite $\epsilon=\frac{1-\sqrt{1-4p}}{2}$. The single photon pair can be prepared in an entangled state in various degrees of freedom, such as polarization and time-bin \cite{yuan2010entangled}. One photon of the entangled pair from each source is sent through a quantum channel for the joint Bell state measurement (BSM). Conditioned on the outcome of the BSM, the other two photons---one from each source---are projected into an entangled state. However, for a LO-BSM which interferes two identical photons at a beamsplitter, two pairs from one source and no pairs produced from the other source will induce an apparent successful outcome of the BSM, without resulting in the desired entanglement swapping. Moreover, the probability of two pairs from one source and one pair from each source is the same, given two sources with same efficiencies. As a result, without postselection on the presence of exactly one photon in each of the other modes of the two sources, the fidelity of entanglement swapping is dramatically reduced. To calculate the fidelity of entanglement swapping in the presence of multiphoton emission and channel losses, we do not specify how the entangled photon pair is prepared or how the LO-BSM is implemented. Instead, we only consider the probability of  photons arriving at the LO-BSM given the SPDC source described by Eq. \ref{eq:SPDC}. The scaling of the resulting fidelity in terms of single-photon-pair probability and channel loss should hold for general cases. As such, suppose the state of photons arriving at the BSM is given by the density matrix $\rho\propto|\alpha|^2|1_A1_B\rangle\langle1_A1_B|+|\beta|^2|\phi\rangle\langle\phi|$, where $|\phi\rangle$ represents states other than $|1_A1_B\rangle$, the fidelity is calculated by $\mathcal{F}=|\alpha|^2/(|\alpha|^2+|\beta|^2)$ \cite{kok2000postselected}. 

To gain some insight of the impact of multiphoton emissions on the non-postselected fidelity, we first consider only the leading-order effect of two input photons to the BSM. The fidelity of entanglement swapping, in the case of lossless channels, is given by 
\be
		\mathcal{F} =\frac{p_{A}p_{B}}{p_{A}p_{B}+p_{A}^2+p_{B}^2}\leq \frac{1}{3},
	\label{eq:swapLOF}
\ee
where we have assumed weak-driving SPDCs and thus $p_{A(B)}^2$ is the two-pair probability of source $A(B)$. The fidelity is bounded by $1/3$ and the bound is saturated for $p_{A}=p_{B}$. When one of the input channels of the BSM (e.g., from source $B$) has loss $\eta$, the fidelity becomes
\be\label{loeswapest}
\mathcal{F} =\frac{\eta p_{A}p_{B}}{\eta p_{A}p_{B}+p_{A}^2+\eta ^2p_{B}^2}\leq \frac{1}{3}
\ee
and the upper bound is saturated for $p_{A}=\eta p_{B}$. This means the source in the lossless channel must be attenuated to match the photon flux of the lossy channel in order to achieve the optimal fidelity, which leads to reduced entangling rate. If the two sources are driven at the same rate, i.e., $p_{A}=p_{B}$, then the fidelity is substantially reduced to $\mathcal{F} \approx \eta$.

Now we proceed to the calculation of fidelity for the general case with two lossy channels and include the contribution of arbitrary multiphoton emissions. We introduce $P(k|n,l|m)$, which denotes the probability that source $A(B)$ emits $n(m)$ pairs of photons, of which $k(l)$ photons reach the BSM. Using Eq. \ref{eq:SPDC}, $P(k|n,l|m)$ is given by
\begin{equation}
		P(k|n,l|m) = (1-\epsilon_A)(1-\epsilon_B)\epsilon_A^n\epsilon_B^mC_n^k\eta_A^k(1-\eta_A)^{n-k}C_m^l\eta_B^l(1-\eta_B)^{m-l}.
\end{equation}
The probability of the entanglement-heralding photon event is $P(1|1,1|1)$, i.e., when both sources emit one pair and both photons reach the BSM. 
The total BSM probability, though, is $1-P_0-P_1$, where $P_0$ and $P_1$ is the probability of zero and one photon that reaches the BSM, respectively:
\be
 P_0=\sum\limits_{n,m\geq0}P(0|n,0|m),
\ee
\be
P_1=\sum\limits_{n\geq1,m\geq0}P(1|n,0|m)+\sum\limits_{n\geq0,m\geq1}P(0|n,1|m).
\ee
As a result, the fidelity is found to be (see Appendix \ref{App:calculation})
\bqa\label{LOF}
\nonumber \mathcal{F} &=&\frac{P(1|1,1|1)}{1-P_0-P_1}\\
&=& \frac{(1-\epsilon_A)(1-\epsilon_B)\epsilon_A\epsilon_B\eta_A\eta_B}{1-\frac{1-\epsilon_A}{1-\epsilon_A(1-\eta_A)}\frac{1-\epsilon_B}{1-\epsilon_B(1-\eta_B)}\left(1+\frac{\epsilon_A\eta_A}{1-\epsilon_A(1-\eta_A)} +\frac{\epsilon_B\eta_B}{1-\epsilon_B(1-\eta_B)} \right)}.
\eqa
It can be shown that
\be\label{LOFbound}
\mathcal{F} \leq \frac{1}{3}\big(1-\epsilon_A(1-\eta_A)\big)^2\big(1-\epsilon_B(1-\eta_B)\big)^2,
\ee
where the upper bound is achieved when $\eta_A\epsilon_A/(1-\epsilon_A)=\eta_B\epsilon_B/(1-\epsilon_B)$. Obviously, $\mathcal{F}\leq 1/3$, which is consistent with the result obtained above by considering the leading-order effect.

\subsection{Balanced loss}

If the loss of the two input channels of the BSM is the same, $\eta_A=\eta_B=\eta$, and the pair generation efficiency of the two sources is the same, $\epsilon_A=\epsilon_B=\epsilon$, Eq. \ref{LOF} reduces to
\be
\mathcal{F}=\frac{(1-\epsilon)^2(1-\epsilon+\epsilon\eta)^3}{3(1-\epsilon)+\epsilon\eta}.
\ee
For $\eta\ll 1$, 
\be\label{fbl}
\mathcal{F}\approx \frac{1}{3}(1-\epsilon)^4=\frac{1}{3}\left(\frac{1+\sqrt{1-4p}}{2}\right)^4.
\ee

\subsection{Unbalanced loss}

If the loss of the two input channels of the BSM is unbalanced with, e.g.,  $\eta_B\ll\eta_A\ll1$, then Eq. \ref{LOFbound} reduces to
\be\label{fubl}
\mathcal{F}\lesssim \frac{1}{3}(1-\epsilon_B)^2=\frac{1}{3}\left(\frac{1+\sqrt{1-4p_B}}{2}\right)^2,
\ee
where the maximum is achieved when $\epsilon_A=\frac{\epsilon_B}{1-\epsilon_B}\frac{\eta_B}{\eta_A}$ or $p_A\approx \frac{\eta_B}{\eta_A}p_B$. This means source $A$ needs to be attenuated such that the photon fluxes of the two input channels of the BSM are equalized.

\section{Nonlinear-optical Bell state measurement}

To illustrate entanglement swapping heralded by sum-frequency generation (Fig. \ref{fig:eswapping}b), we consider two SPDC sources generating time-bin entangled photon pairs. Suppose the joint state of the four photons is $\ket{\Phi^+}_{12}\otimes \ket{\Phi^+}_{34}=\frac{1}{2}(\ket e_1\ket e_2+\ket l_1\ket l_2)\otimes (\ket e_3\ket e_4+\ket l_3\ket l_4)$, where $e$ and $l$ represent early and late time bins, respectively. Photons 2 and 3 are subject to SFG. Conditioned on the SFG, which occurs with a small probability, the joint photon state becomes
\bqa\label{sfgteleport}
\nonumber\ket{\Phi^+}_{12}\otimes \ket{\Phi^+}_{34}&\xrightarrow{\text{SFG}}&\frac{1}{2}(\ket{e}_\Sigma\ket{e}_1\ket{e}_4+\ket{l}_\Sigma\ket{l}_1\ket{l}_4)\\
&=&\frac{1}{2\sqrt{2}}\left(\ket{\Sigma^{+}}(\ket{e}_1\ket{e}_4+\ket{l}_1\ket{l}_4)+\ket{\Sigma^{-}}(\ket{e}_1\ket{e}_4-\ket{l}_1\ket{l}_4)\right)
\eqa
where $\ket{\Sigma^{\pm}}=\frac{1}{\sqrt{2}}( \ket e_{\Sigma}\pm \ket l_{\Sigma})$ are two orthogonal SFG photon states. As a result, conditioned on the detection of the SFG photon in the states $\ket{\Sigma^{\pm}}$, i.e., the NLO-BSM, the joint state of photons 1 and 4 becomes $\ket{\Phi^+}_{14}=\frac{1}{\sqrt{2}}(\ket e_1\ket e_4+\ket l_1\ket l_4)$ or $\ket{\Phi^-}_{14}=\frac{1}{\sqrt{2}}(\ket e_1\ket e_4-\ket l_1\ket l_4)$. Here we have assumed a single SFG element to induce interactions between the early time bins $\ket e_1,\ket e_4$ or late time bins $\ket l_1,\ket l_4$. Introducing a second SFG element to interact $\ket e$ and $\ket l$ will result in entangled photons 1 and 4 in any of the four Bell states, depending on the outcome of the NLO-BSM of the two SFG elements, realizing a complete Bell state measurement \cite{kim2001quantum} (see Appendix \ref{App:completeBSM}). While here the entanglement swapping via SFG is illustrated using time-bin entanglement, it can be generalized to other degrees of freedom of photons \cite{kim2001quantum,tsujimoto2024experimental}. 

To calculate the fidelity of the SFG-heralded entanglement swapping in the presence of lossy channels, we again do not assume the form of the initial entangled photons and the implementation of the projection measurement of the SFG photon. Suppose the SFG process involves three modes $a$, $b$, and $c$, and is described by the Hamiltonian $H/\hbar = g\left(abc^\dagger + h.c.\right)$. The evolution of an initial state $\ket{n_an_b0}$ is given by
\begin{equation}
		e^{-iHt/\hbar}\ket{n_an_b0} \approx (1-iHt/\hbar)\ket{n_an_b0} =\ket{n_an_b0}-i \sqrt{p_\mathrm{SFG}n_an_b} \vert (n_a-1) (n_b-1) 1 \rangle),
	\label{eq:SFG}
\end{equation}
where we have assumed a weak SFG process with single-photon SFG probability $p_\mathrm{SFG}=(gt)^2\ll 1$. This assumption is used for the fidelity calculation below for arbitrary channel loss and source efficiency. In the case of large channel loss or low source efficiency, the same result is obtained for arbitrary SFG efficiency. We note difference-frequency generation (DFG), i.e., generating a mode-$b$ photon from a pair of photons in modes $a$ and $c$, cannot be used for NLO-BSM because of the spontaneous down-conversion of a mode-$c$ photon into a pair of photons in modes $a$ and $b$, which has a probability similar to the DFG process.  

In general, the probability for generating a SFG photon given source $A(B)$ emits $n(m)$ pairs of photons of which $k(l)$ photons reach the NLO-BSM is given by
\begin{equation}
		\bar P(k|n,l|m) = (1-\epsilon_A)(1-\epsilon_B)\epsilon_A^n\epsilon_B^mC_n^k\eta_A^k(1-\eta_A)^{n-k}C_m^l\eta_B^l(1-\eta_B)^{m-l}klp_{\mathrm{SFG}}.
\end{equation}
The probability for the faithful entanglement swapping is  
\begin{equation}
		\bar P(1|1,1|1) = (1-\epsilon_A)(1-\epsilon_B)\epsilon_A\epsilon_B\eta_A\eta_Bp_{\mathrm{SFG}}.
\end{equation}
The total probability for SFG photon generation is
\bqa
\nonumber&&\sum\limits_{\substack{n,m\geq0\\ 0\leq k\leq n\\ 0\leq l\leq m}}\bar P(k|n,l|m) \\
\nonumber&=&p_{\mathrm{SFG}}(1-\epsilon_A)(1-\epsilon_B)\sum\limits_{n,m\geq0} \epsilon_A^n\epsilon_B^m n\eta_A m\eta_B\\
&=&p_{\mathrm{SFG}}\eta_A \eta_B\frac{\epsilon_A}{1-\epsilon_A}\frac{\epsilon_B}{1-\epsilon_B}.
\eqa
As a result, the fidelity for SFG-heralded entanglement swapping is 
\bqa\label{fsfg}
		\nonumber\mathcal{F} &=& \frac{\bar P(1|1,1|1)}{\sum\limits_{\substack{n,m\geq0\\ 0\leq k\leq n\\ 0\leq l\leq m}}\bar P(k|n,l|m) }\\
		\nonumber&=&(1-\epsilon_A)^2(1-\epsilon_B)^2\\
		&=&\left(\frac{1+\sqrt{1-4p_{A}}}{2}\right)^2\left(\frac{1+\sqrt{1-4p_{B}}}{2}\right)^2.
\eqa
In contrast to the LO-BSM approach, the fidelity of the NLO-BSM heralded entanglement swapping is independent of channel losses. One can show that this result also applies to the case of efficient SFG with relatively large $p_{\mathrm{SFG}}$ and with two input photons to the NLO-BSM, i.e., for lossy channels with $\eta_A,\ \eta_B\ll 1$ or inefficient entanglement sources where we only have to consider the leading-order photon events.

\begin{figure}[!htb]
\begin{center}
\includegraphics[width=0.5\columnwidth]{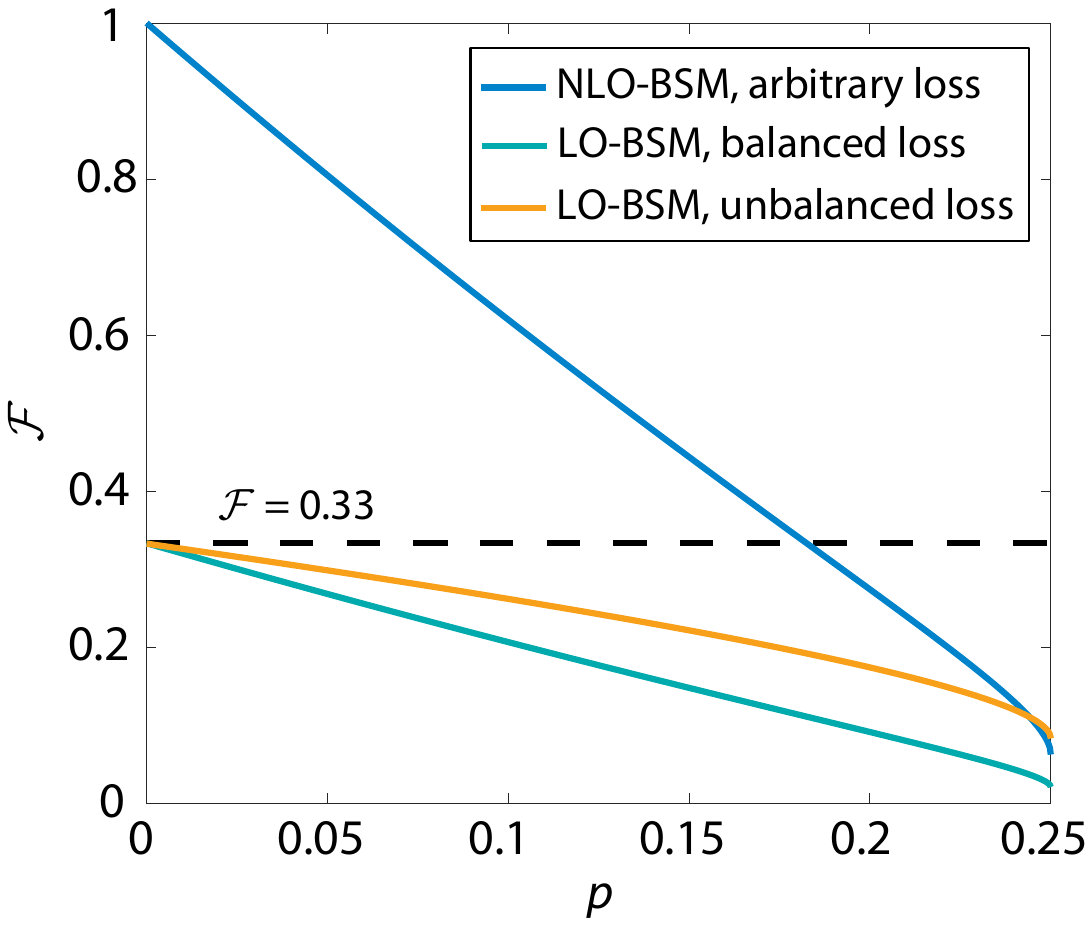}
\caption{\textbf{Non-postselected fidelity of LO-BSM and NLO-BSM heralded entanglement swapping.} The NLO-BSM result applies to arbitrary loss conditions and is plotted for $p_A=p_B=p$. For LO-BSM with balanced loss, we assume $\eta_A=\eta_B\ll 1$ and $p_A=p_B=p$. For LO-BSM with unbalanced loss, we assume $\eta_B\ll\eta_A\ll 1$ and $p_A=\eta_Bp_B/\eta_A\ (p_B=p)$, which corresponds to the optimal fidelity. Dashed line indicates the general bound of $\mathcal{F}=1/3$ for the LO-BSM approach. }
\label{fig:fidelity}
\end{center}
\end{figure}

In Fig. \ref{fig:fidelity}, we compare the fidelity of NLO-BSM and LO-BSM heralded entanglement swapping for two practical loss conditions: balanced loss ($\eta_A=\eta_B\ll1$) and unbalanced loss ($\eta_B\ll\eta_A\ll1$). The fidelity of the LO-BSM approach is calculated using Eq. \ref{fbl} and Eq. \ref{fubl} for the balanced loss and unbalanced loss, respectively, while the fidelity of the NLO-BSM approach, which is universal, is calculated using Eq. \ref{fsfg}. The fidelity of the NLO-BSM is consistently higher than that of the LO-BSM for all pair-generation rate. 

Lastly, we comment on the benefit of a NLO-BSM in relaxing the strict requirement for indistinguishable photons for a LO-BSM. The NLO-BSM requires two photons of different frequencies in the first place, so the discussion here is regarding the frequency indistinguishability of successive photons in the same mode. For a NLO-BSM implemented for polarization qubits, the projection of the SFG photon is in the polarization space, so frequency variations of input photons will not affect the fidelity (the efficiency will be reduced if the variation is greater than the bandwidth of the SFG). For a NLO-BSM implemented for time-bin qubits, the BSM requires projection of the time-bin SFG photon via an unbalanced interferometer. If the frequency variation of the input photons is comparable to the inverse of the time between early and late time bins, the relative phase between the early and late time bins of the projected state will vary. And if the resulting phase variation is unknown or cannot be compensated, the fidelity of the swapped entanglement will be reduced.

\section{Nanophotonics-enhanced sum-frequency generation and entangling rate}
Demonstrated single-photon SFG probability of bulk crystals and waveguides is very small, on the order of $10^{-8}$ \cite{kim2001quantum,sephton2023quantum,qiu2023remote,tanzilli2005photonic,sangouard2011faithful,fisher2021single,guerreiro2013interaction,guerreiro2014nonlinear,tsujimoto2024experimental}, due to the limited $\chi^{(2)}$ nonlinearity and weak light confinement. Recently, the development of integrated nonlinear photonics platforms based on strong nonlinear materials, including periodically-poled thin-film LiNbO$_3$ \cite{lu2020toward} and InGaP \cite{zhao2022ingap}, has led to orders-of-magnitude enhancement of the SFG efficiency \cite{akin2025faithful}. In this section, we derive the single-photon SFG probability for cavities and waveguides, and estimate the nanophotonics-enhanced SFG probability and entangling rate. We note, in general, integrated photonic devices are mode-polarization dependent though. However, a Sagnac interferometer-type design can be implemented for the NLO-BSM to be compatible with polarization qubits \cite{kim2006phase,tsujimoto2024experimental}.

\subsection{Cavity SFG probability}
For a triply-resonant $\chi^{(2)}$ cavity with phase-matched $a$, $b$, $c$ modes, the Hamiltonian describing the tri-modal interaction is given by 
\be
H/\hbar=\omega_a a^\dagger a+\omega_b b^\dagger b+\omega_c c^\dagger c+g(a^\dagger b^\dagger c+abc^\dagger).
\ee
The coupled-mode equations of the cavity modes are given by
\bqa
\frac{da}{dt}=-(i\omega_a+\frac{\kappa_a}{2})a-igb^\dagger c+i\sqrt{\frac{\kappa_{ae}}{2}}a_{\mathrm{in}},\\
\frac{db}{dt}=-(i\omega_b+\frac{\kappa_b}{2})b-iga^\dagger c+i\sqrt{\frac{\kappa_{be}}{2}}b_{\mathrm{in}},\\
\frac{dc}{dt}=-(i\omega_c+\frac{\kappa_c}{2})c-igab+i\sqrt{\frac{\kappa_{ce}}{2}}c_{\mathrm{in}},
\eqa
where $\kappa_k$ and $\kappa_{ke}$ are the total and external dissipation rate of mode $k$, respectively. 

For the SFG process with continuously-pumped modes $a$ and $b$, the static cavity amplitude of the three modes, to the leading order of $g/\kappa$, are found by solving the coupled-mode equations:
\bqa
a=\frac{i\sqrt{\frac{\kappa_{ae}}{2}}}{i(\omega_a-\omega_{pa})+\frac{\kappa_a}{2}}a_{\mathrm{in}},\\
b=\frac{i\sqrt{\frac{\kappa_{be}}{2}}}{i(\omega_b-\omega_{pb})+\frac{\kappa_b}{2}}b_{\mathrm{in}},\\
c=-\frac{igab}{i(\omega_c-\omega_{pa}-\omega_{pb})+\frac{\kappa_c}{2}},
\eqa
where $\omega_{pa}$ and $\omega_{pb}$ are the pump frequencies of mode $a$ and $b$, respectively. From this, we obtain the relation between the outgoing SFG power and incoming pump powers:
\be\label{etasfgdef}
P_c=\eta_{\mathrm{SFG}}P_aP_b,
\ee
where $P_c=\frac{\kappa_{ce}}{2}\hbar\omega_c|c|^2$, $P_a=\hbar\omega_a|a_{\mathrm{in}}|^2$, $P_b=\hbar\omega_b|b_{\mathrm{in}}|^2$, with the SFG efficiency given by 
\be
\eta_{\mathrm{SFG}}=g^2\frac{\kappa_{ae}/2}{(\omega_a-\omega_{pa})^2+(\kappa_{a}/2)^2}\frac{\kappa_{be}/2}{(\omega_b-\omega_{pb})^2+(\kappa_{b}/2)^2}\frac{\kappa_{ce}/2}{(\omega_c-\omega_{pa}-\omega_{pb})^2+(\kappa_{c}/2)^2}\frac{\hbar\omega_c}{\hbar\omega_a\hbar\omega_b}.
\ee
In the case of on-resonance driving, $\omega_{pa(b)}=\omega_{a(b)}$, and  frequency-matching condition, $\omega_c=\omega_a+\omega_b$, 
\be\label{etasfg}
\eta_{\mathrm{SFG}}=g^2\frac{\kappa_{ae}/2}{(\kappa_{a}/2)^2}\frac{\kappa_{be}/2}{(\kappa_{b}/2)^2}\frac{\kappa_{ce}/2}{(\kappa_{c}/2)^2}\frac{\hbar\omega_c}{\hbar\omega_a\hbar\omega_b}.
\ee
Eq. \ref{etasfgdef} relates to the characterization of a SFG cavity: two independent beams of power $P_a$ and $P_b$ are used to drive the cavity and generate the SFG beam.

Next we derive the inherent single-photon SFG probability $p_{\mathrm{SFG}}$ of the nonlinear cavity, which can be defined via
\be\label{psfgdef}
n_c\kappa_c=p_{\mathrm{SFG}}n_an_b\kappa_a.
\ee
The L.H.S. represents the total SFG photon flux and the R.H.S. represents the flux of the cluster of $a$ and $b$ photons. We have assumed $\kappa_a\approx \kappa_b$.  Using
\be
n_a=\frac{\kappa_{ae}/2}{(\kappa_a/2)^2}\frac{P_a}{\hbar\omega_a}, \quad
n_b=\frac{\kappa_{be}/2}{(\kappa_b/2)^2}\frac{P_b}{\hbar\omega_b}, \quad
P_c=\hbar\omega_cn_c\frac{\kappa_{ce}}{2},
\ee
and the definitions of Eqs. \ref{etasfgdef} and \ref{psfgdef}, we find
\be\label{relation}
p_{\mathrm{SFG}}=\eta_{\mathrm{SFG}}\frac{(\kappa_{a}/2)^2}{\kappa_{ae}/2}\frac{(\kappa_{b}/2)^2}{\kappa_{be}/2}\frac{\kappa_{c}}{\kappa_a\kappa_{ce}/2}
\frac{\hbar\omega_a\hbar\omega_b}{\hbar\omega_c}.
\ee
Finally, using Eq. \ref{etasfg}, we obtain
\be\label{psfg}
p_{\mathrm{SFG}}=\frac{4g^2}{\kappa_{a}\kappa_{c}}.
\ee
This result applies to photons with a bandwidth up to the cavity linewidth. 

State-of-the-art integrated nonlinear photonics platforms enable $p_{\mathrm{SFG}}$ orders-of-magnitude higher than bulk crystals and waveguides. For example, for the recently developed InGaP platform \cite{zhao2022ingap,akin2024ingap}, a 5-$\mu$m-radius microring with $Q_{1550\mathrm{nm}}\approx 4\times 10^5$, $Q_{775\mathrm{nm}}\approx 10^5$ and $g/2\pi\approx 20$ MHz can be achieved (note $g$ of SFG is twice that of second-harmonic generation (SHG) based on the induced polarization \cite{boyd2020nonlinear}), leading to $p_{\mathrm{SFG}}\approx  10^{-3}$. Experimentally, $p_{\mathrm{SFG}}=4\times 10^{-5}$ has been realized using a 10-$\mu$m-radius InGaP microring \cite{akin2025faithful}. For periodically-poled LiNbO$_3$, based on the measured $Q_{1550\mathrm{nm}}$, $Q_{775\mathrm{nm}}$, and $g$ of SHG \cite{lu2020toward}, the estimated $p_{\mathrm{SFG}}$ is about $10^{-4}$.

\subsection{Waveguide SFG probability}
For SFG in a waveguide, the SFG probability is given by \cite{sangouard2011faithful}
\be
p_{\mathrm{SFG}}=\eta_{\mathrm{SFG}}\bar P L^2=2\pi\eta_{\mathrm{SFG}}h\nu\Delta\nu L^2,
\ee
where $\bar P$, $\nu$ and $\Delta\nu$ is the average power, frequency and bandwidth of the single-photon pulse, respectively, $L$ is the waveguide length, and $\eta_{\mathrm{SFG}}$ is the SFG efficiency defined in Eq. \ref{etasfgdef}. 
If we utilize the full bandwidth of the waveguide, i.e., $\Delta\nu=\hat\Delta\nu/L$, where $\hat\Delta\nu$ is the spectral acceptance of the waveguide \cite{sangouard2011faithful}, then 
\be
p_{\mathrm{SFG}}=2\pi\eta_{\mathrm{SFG}}h\nu\hat\Delta\nu L.
\ee
As an example, for phase-matched InGaP nanophotonic waveguides, $\eta_{\mathrm{SFG}}=4\eta_{\mathrm{SHG}}=500,000\%$/W/cm$^2$ and $\hat\Delta\nu=6$ GHz$\cdot$cm \cite{akin2024ingap}. Thus, for a $L=1$ cm long waveguide, we estimate $p_{\mathrm{SFG}}\approx3\times 10^{-5}$.

\subsection{Entanglement swapping rate}

With the state-of-the-art nanophotonics-enhanced SFG efficiency, the NLO-BSM-based entanglement swapping can actually be more efficient than the LO-BSM counterpart in some cases, including the loss-unbalanced case. The entanglement rate for the LO-BSM approach in this case is given by 
\be\label{LOR}
R_{\mathrm{LO}}=\eta_A\eta_Bp_{A} p_{B}R_c=\eta_B^2 p_{B}^2R_c,
\ee
where we have used $p_{A}=\eta_B p_{B}/\eta_A$ for the optimal fidelity and $R_c$ is the clock rate. Source $A$ needs to be substantially attenuated to achieve the optimal fidelity, which  nevertheless is less than 1/3. In contrast, the entanglement rate for NLO-BSM is given by
\be\label{NLOR}
R_{\mathrm{NLO}}=p_{\mathrm{SFG}}\eta_A\eta_Bp_{A} p_{B}R_c=p_{\mathrm{SFG}}\eta_A\eta_B p_{B}^2R_c,
\ee
where $p_{A}$ can be as large as $p_{B}$. Comparing Eq. \ref{NLOR} and Eq. \ref{LOR}, if $p_{\mathrm{SFG}}>\eta_B/\eta_A$, the NLO-BSM approach can be more efficient than the LO-BSM approach. This is relevant, for example, for satellite-mediated entanglement swapping, given the available $p_{\mathrm{SFG}}=10^{-4}-10^{-3}$ using the state-of-the-art nanophotonics platforms while the loss of satellite-ground link can be $>50$ dB \cite{ren2017ground}.

If the fidelity for the NLO-BSM approach were to be lowered and be the same as the LO-BSM approach, then the entanglement source can be driven more efficiently and the entanglement rate will be increased for the NLO-BSM approach. For example, if the targeted fidelity is 1/3, then the single-photon probability can be about 0.2 for the NLO-BSM approach (see Fig. \ref{fig:fidelity}), while that of the LO-BSM approach will be at least one order of magnitude smaller. As a result, the two-photon probability $p_Ap_B$ for the NLO-BSM approach will be 100 times higher than the LO-BSM counterpart, and with $p_{\mathrm{SFG}}=10^{-4}-10^{-3}$, the entanglement rate is only 10 to 100 times lower than the LO-BSM approach under general loss conditions. On the other hand, NLO-BSM avoids the stringent requirement of identical photons, which will significantly ameliorate the technical challenges in implementation of quantum networking protocols for real-world applications. 

Alternatively, we note that LO-BSM can be used in combination with heralded entanglement generation \cite{sliwa2003conditional,barz2010heralded,wagenknecht2010experimental} to perform high fidelity non-postselected entanglement swapping. But this comes with a substantial reduction in rate. For example, for the six-photon scheme \cite{sliwa2003conditional}, each heralded entanglement source needs to generate two ancillary photon pairs for every entangled pair needed. Thus, even at the demonstrated $4 \times 10^{-5}$ efficiency, a NLO-BSM-based scheme can operate at a much higher rate than LO schemes based on heralded entanglement sources for the same fidelity \cite{sangouard2011faithful}. 

\section{Conclusion}

We analyzed the fidelity of LO-BSM and NLO-BSM heralded entanglement swapping in lossy channels and derived general results for arbitrary loss conditions. Compared to LO-BSM, the NLO-BSM-based approach can achieve higher fidelity with the same photon source rate or operate at a much higher photon source rate while maintaining the same fidelity. Additionally, with the enhanced SFG efficiency provided by advanced nonlinear nanophotonics and the relaxation of the identical-photon requirement, NLO-BSM heralded entanglement swapping shows great promise for practical real-world applications.

\appendix
\section{Calculation of the fidelity of LO-BSM entanglement swapping}\label{App:calculation}
The probability of zero and one photon that reaches the LO-BSM, $P_0$ and $P_1$, can be calculated as:
\bqa
\nonumber P_0&=&\sum\limits_{n,m\geq0}P(0|n,0|m)\\
 \nonumber &=& \sum\limits_{n,m\geq0}(1-\epsilon_A)(1-\epsilon_B)\epsilon_A^n\epsilon_B^m(1-\eta_A)^{n}(1-\eta_B)^{m}\\
 &=&(1-\epsilon_A)(1-\epsilon_B)\frac{1}{1-\epsilon_A(1-\eta_A)}\frac{1}{1-\epsilon_B(1-\eta_B)},
\eqa
\bqa
\nonumber P_1&=&\sum\limits_{n\geq1,m\geq0}P(1|n,0|m)+\sum\limits_{n\geq0,m\geq1}P(0|n,1|m)\\
 \nonumber &=& (1-\epsilon_A)(1-\epsilon_B)\left(\sum\limits_{n\geq1,m\geq0}\epsilon_A^n\epsilon_B^mC_n^1\eta_A(1-\eta_A)^{n-1}(1-\eta_B)^{m}+\sum\limits_{n\geq0,m\geq1}\epsilon_A^n\epsilon_B^mC_m^1(1-\eta_A)^{n}\eta_B(1-\eta_B)^{m-1}\right)\\
 &=&(1-\epsilon_A)(1-\epsilon_B)\left(\epsilon_A\eta_A\frac{1}{\big(1-\epsilon_A(1-\eta_A)\big)^2}\frac{1}{1-\epsilon_B(1-\eta_B)}+\epsilon_B\eta_B\frac{1}{1-\epsilon_A(1-\eta_A)}\frac{1}{\big(1-\epsilon_B(1-\eta_B)\big)^2}\right).
\eqa

As a result, the fidelity is given by
\bqa\label{app:LOF}
\nonumber \mathcal{F}&=&\frac{P(1|1,1|1)}{1-P_0-P_1}\\
&=& \frac{(1-\epsilon_A)(1-\epsilon_B)\epsilon_A\epsilon_B\eta_A\eta_B}{1-\frac{1-\epsilon_A}{1-\epsilon_A(1-\eta_A)}\frac{1-\epsilon_B}{1-\epsilon_B(1-\eta_B)}\left(1+\frac{\epsilon_A\eta_A}{1-\epsilon_A(1-\eta_A)} +\frac{\epsilon_B\eta_B}{1-\epsilon_B(1-\eta_B)} \right)}.
\eqa

Re-writing Eq. \ref{app:LOF} allows to find the upper bound of the fidelity:
\bqa
\nonumber \mathcal{F}&=&\frac{(1-\epsilon_A)(1-\epsilon_B)\epsilon_A\epsilon_B\eta_A\eta_B\big(1-\epsilon_A(1-\eta_A)\big)^2\big(1-\epsilon_B(1-\eta_B)\big)^2}{(1-\epsilon_B)^2\epsilon_A^2\eta_A^2+(2\epsilon_A\eta_A+1-\epsilon_A)(1-\epsilon_B)\epsilon_B\eta_B\epsilon_A\eta_A+\epsilon_B^2\eta_B^2(\epsilon_A\eta_A+1-\epsilon_A)^2}\\
\nonumber &\lesssim &\frac{(1-\epsilon_A)(1-\epsilon_B)\epsilon_A\epsilon_B\eta_A\eta_B\big(1-\epsilon_A(1-\eta_A)\big)^2\big(1-\epsilon_B(1-\eta_B)\big)^2}{(1-\epsilon_B)^2\epsilon_A^2\eta_A^2+(1-\epsilon_A)(1-\epsilon_B)\epsilon_B\eta_B\epsilon_A\eta_A+\epsilon_B^2\eta_B^2(1-\epsilon_A)^2}\\
&\leq & \frac{1}{3}\big(1-\epsilon_A(1-\eta_A)\big)^2\big(1-\epsilon_B(1-\eta_B)\big)^2,
\eqa
where the maximum is achieved when $(1-\epsilon_B)\epsilon_A\eta_A=(1-\epsilon_A)\epsilon_B\eta_B$.

\section{Complete Bell state measurement}\label{App:completeBSM}

\begin{figure*}[!htb]
	\begin{center}
		\includegraphics[width=0.7\columnwidth]{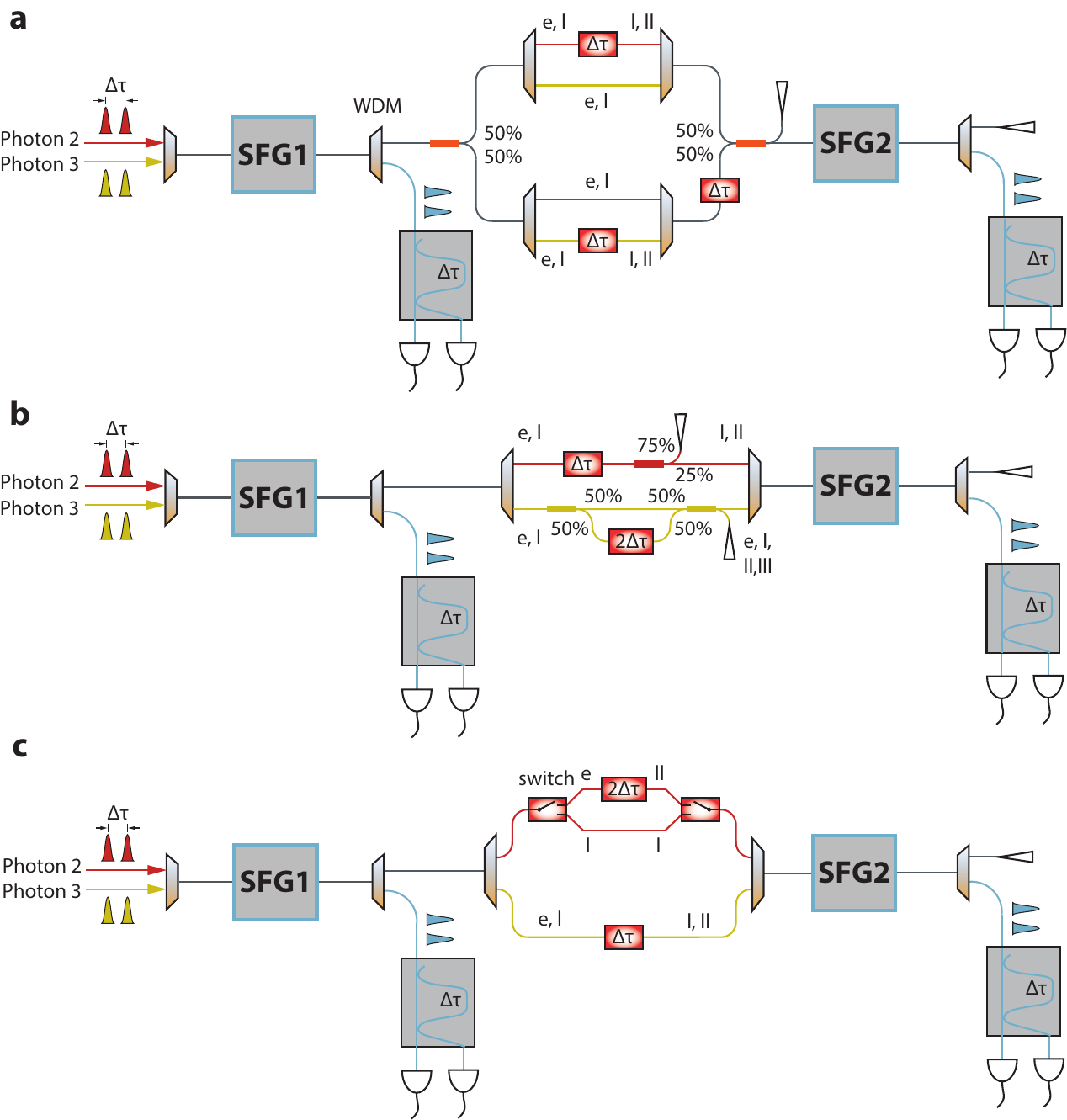}
		\caption{\textbf{Complete Bell state measurement for time-bin encoded photons.} \textbf{a} and \textbf{b}. Two ``lossy''schemes using beamplitters. \textbf{c}.  A ``lossless'' scheme using optical switches. }
		\label{fig:completeBSA}
	\end{center}
\end{figure*}

Complete Bell state measurements can be implemented using two nonlinear SFG elements. To show this for the entanglement swapping, suppose the initial joint state of the four photons is expressed as 
\bqa\label{sfgteleport}
\ket{\Phi^+}_{12}\otimes \ket{\Phi^+}_{34}&=&\frac{1}{2}(\ket e_1\ket e_2+\ket l_1\ket l_2)\otimes (\ket e_3\ket e_4+\ket l_3\ket l_4)\\
&=&\frac{1}{2}\big(\ket{\Phi^+}_{14}\ket{\Phi^+}_{23}+\ket{\Phi^-}_{14}\ket{\Phi^-}_{23}+\ket{\Psi^+}_{14}\ket{\Psi^+}_{23}+\ket{\Psi^-}_{14}\ket{\Psi^-}_{23}\big),
\eqa
where $\ket{\Phi^\pm}_{23}=\frac{1}{\sqrt{2}}(\ket e_2\ket e_3\pm\ket l_2\ket l_3)$ and  $\ket{\Psi^\pm}_{23}=\frac{1}{\sqrt{2}}(\ket e_2\ket l_3\pm\ket l_2\ket e_3)$ are the four Bell states of photons 2 and 3. To convert the four Bell states to four distinct SFG single-photon states, two SFG elements are required. One SFG element will interact $e(l)$ with $e(l)$ and convert $\ket{\Phi^\pm}_{23}$ to two orthogonal SFG states, while the second SFG element will only interact $e$ with $l$ and convert $\ket{\Psi^\pm}_{23}$ to two orthogonal SFG states in different spatial modes. As a result, conditioned on the SFG, the joint state of the initial four photons becomes 
\be
\ket{\Phi^+}_{12}\otimes \ket{\Phi^+}_{34}\xrightarrow{\text{SFG}}\frac{1}{2}\big(\ket{\Sigma_1^+} \ket{\Phi^+}_{14} +\ket{\Sigma_1^-}\ket{\Phi^-}_{14}+\ket{\Sigma_2^+} \ket{\Psi^+}_{14} +\ket{\Sigma_2^-}\ket{\Psi^-}_{14} \big),
\ee
where $\ket{\Sigma_{1(2)}^{\pm}}=\frac{1}{\sqrt{2}}( \ket e_{\Sigma_{1(2)}}\pm \ket l_{\Sigma_{1(2)}})$ are the orthogonal SFG photon states generated by the first and second nonlinear elements. The four SFG states distinguish the four Bell states completely. By measuring the SFG photon in one of the states, the joint state of photons 1 and 4 is projected to one of the four Bell states. 

Two schemes of the complete Bell state measurement are illustrated in Fig. \ref{fig:completeBSA}. For the first scheme (Fig. \ref{fig:completeBSA}a), the first nonlinear element induces interaction between $ee$ or $ll$ time bins to generate $e$ and $l$ SFG photons, whose amplitudes interfere at an unbalanced MZI for projection measurement. Before the second nonlinear cavity, the two photons pass through two sets of wavelength-selective (via WDMs) delay lines to delay the two photons and temporally align $\ket e_2\ket l_3$ and $\ket l_2\ket e_3$, respectively. $\ket e_2\ket e_3$ and $\ket l_2\ket l_3$ will be temporally mismatched after the delay lines. The second nonlinear element then induces interaction only between the $e$ and $l$ time bins, and the generated $e$ and $l$ SFG photons (because of the relative delay between the two sets of delay lines) interfere at an unbalanced MZI for projection measurement. In the second scheme (Fig. \ref{fig:completeBSA}b), after the same first SFG element, wavelength-selective delay lines are introduced to convert photon 2's bins $e\rightarrow l$, $l\rightarrow ll$ and photon 3's bins $e\rightarrow (e,ll)$,  $l\rightarrow (l,lll)$. Thus, only $\ket e_2\ket l_3$ and $\ket l_2\ket e_3$ become temporally aligned and can generate SFG photons in the second nonlinear element. Using optical switches in replacement of beamsplitters, the nonlinear Bell state analyzer can be made ``lossless'' (aside from the finite SFG efficiency), as shown in Fig. \ref{fig:completeBSA}c as an example.

\vspace{2mm}
\noindent\textbf{Acknowledgements}\\ 
We thank Paul G. Kwiat for the discussion. This work is supported by US National Science Foundation under Grant No. 2223192 and QLCI-HQAN (Grant No. 2016136) and U.S. Department of Energy Office of Science National Quantum Information Science Research Centers. 


\begin{thebibliography}{31}%
\makeatletter
\providecommand \@ifxundefined [1]{%
 \@ifx{#1\undefined}
}%
\providecommand \@ifnum [1]{%
 \ifnum #1\expandafter \@firstoftwo
 \else \expandafter \@secondoftwo
 \fi
}%
\providecommand \@ifx [1]{%
 \ifx #1\expandafter \@firstoftwo
 \else \expandafter \@secondoftwo
 \fi
}%
\providecommand \natexlab [1]{#1}%
\providecommand \enquote  [1]{``#1''}%
\providecommand \bibnamefont  [1]{#1}%
\providecommand \bibfnamefont [1]{#1}%
\providecommand \citenamefont [1]{#1}%
\providecommand \href@noop [0]{\@secondoftwo}%
\providecommand \href [0]{\begingroup \@sanitize@url \@href}%
\providecommand \@href[1]{\@@startlink{#1}\@@href}%
\providecommand \@@href[1]{\endgroup#1\@@endlink}%
\providecommand \@sanitize@url [0]{\catcode `\\12\catcode `\$12\catcode
  `\&12\catcode `\#12\catcode `\^12\catcode `\_12\catcode `\%12\relax}%
\providecommand \@@startlink[1]{}%
\providecommand \@@endlink[0]{}%
\providecommand \url  [0]{\begingroup\@sanitize@url \@url }%
\providecommand \@url [1]{\endgroup\@href {#1}{\urlprefix }}%
\providecommand \urlprefix  [0]{URL }%
\providecommand \Eprint [0]{\href }%
\providecommand \doibase [0]{http://dx.doi.org/}%
\providecommand \selectlanguage [0]{\@gobble}%
\providecommand \bibinfo  [0]{\@secondoftwo}%
\providecommand \bibfield  [0]{\@secondoftwo}%
\providecommand \translation [1]{[#1]}%
\providecommand \BibitemOpen [0]{}%
\providecommand \bibitemStop [0]{}%
\providecommand \bibitemNoStop [0]{.\EOS\space}%
\providecommand \EOS [0]{\spacefactor3000\relax}%
\providecommand \BibitemShut  [1]{\csname bibitem#1\endcsname}%
\let\auto@bib@innerbib\@empty
\bibitem [{\citenamefont {Kok}\ and\ \citenamefont
  {Braunstein}(2000)}]{kok2000postselected}%
  \BibitemOpen
  \bibfield  {author} {\bibinfo {author} {\bibfnamefont {Pieter}\ \bibnamefont
  {Kok}}\ and\ \bibinfo {author} {\bibfnamefont {Samuel~L}\ \bibnamefont
  {Braunstein}},\ }\bibfield  {title} {\enquote {\bibinfo {title} {Postselected
  versus nonpostselected quantum teleportation using parametric
  down-conversion},}\ }\href@noop {} {\bibfield  {journal} {\bibinfo  {journal}
  {Physical Review A}\ }\textbf {\bibinfo {volume} {61}},\ \bibinfo {pages}
  {042304} (\bibinfo {year} {2000})}\BibitemShut {NoStop}%
\bibitem [{\citenamefont {Pan}\ \emph {et~al.}(2003)\citenamefont {Pan},
  \citenamefont {Gasparoni}, \citenamefont {Aspelmeyer}, \citenamefont
  {Jennewein},\ and\ \citenamefont {Zeilinger}}]{pan2003experimental}%
  \BibitemOpen
  \bibfield  {author} {\bibinfo {author} {\bibfnamefont {Jian-Wei}\
  \bibnamefont {Pan}}, \bibinfo {author} {\bibfnamefont {Sara}\ \bibnamefont
  {Gasparoni}}, \bibinfo {author} {\bibfnamefont {Markus}\ \bibnamefont
  {Aspelmeyer}}, \bibinfo {author} {\bibfnamefont {Thomas}\ \bibnamefont
  {Jennewein}}, \ and\ \bibinfo {author} {\bibfnamefont {Anton}\ \bibnamefont
  {Zeilinger}},\ }\bibfield  {title} {\enquote {\bibinfo {title} {Experimental
  realization of freely propagating teleported qubits},}\ }\href@noop {}
  {\bibfield  {journal} {\bibinfo  {journal} {Nature}\ }\textbf {\bibinfo
  {volume} {421}},\ \bibinfo {pages} {721--725} (\bibinfo {year}
  {2003})}\BibitemShut {NoStop}%
\bibitem [{\citenamefont {Sangouard}\ \emph
  {et~al.}(2011{\natexlab{a}})\citenamefont {Sangouard}, \citenamefont {Simon},
  \citenamefont {De~Riedmatten},\ and\ \citenamefont
  {Gisin}}]{sangouard2011quantum}%
  \BibitemOpen
  \bibfield  {author} {\bibinfo {author} {\bibfnamefont {Nicolas}\ \bibnamefont
  {Sangouard}}, \bibinfo {author} {\bibfnamefont {Christoph}\ \bibnamefont
  {Simon}}, \bibinfo {author} {\bibfnamefont {Hugues}\ \bibnamefont
  {De~Riedmatten}}, \ and\ \bibinfo {author} {\bibfnamefont {Nicolas}\
  \bibnamefont {Gisin}},\ }\bibfield  {title} {\enquote {\bibinfo {title}
  {Quantum repeaters based on atomic ensembles and linear optics},}\
  }\href@noop {} {\bibfield  {journal} {\bibinfo  {journal} {Reviews of Modern
  Physics}\ }\textbf {\bibinfo {volume} {83}},\ \bibinfo {pages} {33--80}
  (\bibinfo {year} {2011}{\natexlab{a}})}\BibitemShut {NoStop}%
\bibitem [{\citenamefont {Azuma}\ \emph {et~al.}(2023)\citenamefont {Azuma},
  \citenamefont {Economou}, \citenamefont {Elkouss}, \citenamefont {Hilaire},
  \citenamefont {Jiang}, \citenamefont {Lo},\ and\ \citenamefont
  {Tzitrin}}]{azuma2023quantum}%
  \BibitemOpen
  \bibfield  {author} {\bibinfo {author} {\bibfnamefont {Koji}\ \bibnamefont
  {Azuma}}, \bibinfo {author} {\bibfnamefont {Sophia~E}\ \bibnamefont
  {Economou}}, \bibinfo {author} {\bibfnamefont {David}\ \bibnamefont
  {Elkouss}}, \bibinfo {author} {\bibfnamefont {Paul}\ \bibnamefont {Hilaire}},
  \bibinfo {author} {\bibfnamefont {Liang}\ \bibnamefont {Jiang}}, \bibinfo
  {author} {\bibfnamefont {Hoi-Kwong}\ \bibnamefont {Lo}}, \ and\ \bibinfo
  {author} {\bibfnamefont {Ilan}\ \bibnamefont {Tzitrin}},\ }\bibfield  {title}
  {\enquote {\bibinfo {title} {Quantum repeaters: From quantum networks to the
  quantum internet},}\ }\href@noop {} {\bibfield  {journal} {\bibinfo
  {journal} {Reviews of Modern Physics}\ }\textbf {\bibinfo {volume} {95}},\
  \bibinfo {pages} {045006} (\bibinfo {year} {2023})}\BibitemShut {NoStop}%
\bibitem [{\citenamefont {Sun}\ \emph {et~al.}(2016)\citenamefont {Sun},
  \citenamefont {Mao}, \citenamefont {Chen}, \citenamefont {Zhang},
  \citenamefont {Jiang}, \citenamefont {Zhang}, \citenamefont {Zhang},
  \citenamefont {Miki}, \citenamefont {Yamashita}, \citenamefont {Terai} \emph
  {et~al.}}]{sun2016quantum}%
  \BibitemOpen
  \bibfield  {author} {\bibinfo {author} {\bibfnamefont {Qi-Chao}\ \bibnamefont
  {Sun}}, \bibinfo {author} {\bibfnamefont {Ya-Li}\ \bibnamefont {Mao}},
  \bibinfo {author} {\bibfnamefont {Si-Jing}\ \bibnamefont {Chen}}, \bibinfo
  {author} {\bibfnamefont {Wei}\ \bibnamefont {Zhang}}, \bibinfo {author}
  {\bibfnamefont {Yang-Fan}\ \bibnamefont {Jiang}}, \bibinfo {author}
  {\bibfnamefont {Yan-Bao}\ \bibnamefont {Zhang}}, \bibinfo {author}
  {\bibfnamefont {Wei-Jun}\ \bibnamefont {Zhang}}, \bibinfo {author}
  {\bibfnamefont {Shigehito}\ \bibnamefont {Miki}}, \bibinfo {author}
  {\bibfnamefont {Taro}\ \bibnamefont {Yamashita}}, \bibinfo {author}
  {\bibfnamefont {Hirotaka}\ \bibnamefont {Terai}},  \emph {et~al.},\
  }\bibfield  {title} {\enquote {\bibinfo {title} {Quantum teleportation with
  independent sources and prior entanglement distribution over a network},}\
  }\href@noop {} {\bibfield  {journal} {\bibinfo  {journal} {Nature Photonics}\
  }\textbf {\bibinfo {volume} {10}},\ \bibinfo {pages} {671--675} (\bibinfo
  {year} {2016})}\BibitemShut {NoStop}%
\bibitem [{\citenamefont {Sun}\ \emph {et~al.}(2017)\citenamefont {Sun},
  \citenamefont {Jiang}, \citenamefont {Mao}, \citenamefont {You},
  \citenamefont {Zhang}, \citenamefont {Zhang}, \citenamefont {Jiang},
  \citenamefont {Chen}, \citenamefont {Li}, \citenamefont {Huang} \emph
  {et~al.}}]{sun2017entanglement}%
  \BibitemOpen
  \bibfield  {author} {\bibinfo {author} {\bibfnamefont {Qi-Chao}\ \bibnamefont
  {Sun}}, \bibinfo {author} {\bibfnamefont {Yang-Fan}\ \bibnamefont {Jiang}},
  \bibinfo {author} {\bibfnamefont {Ya-Li}\ \bibnamefont {Mao}}, \bibinfo
  {author} {\bibfnamefont {Li-Xing}\ \bibnamefont {You}}, \bibinfo {author}
  {\bibfnamefont {Wei}\ \bibnamefont {Zhang}}, \bibinfo {author} {\bibfnamefont
  {Wei-Jun}\ \bibnamefont {Zhang}}, \bibinfo {author} {\bibfnamefont {Xiao}\
  \bibnamefont {Jiang}}, \bibinfo {author} {\bibfnamefont {Teng-Yun}\
  \bibnamefont {Chen}}, \bibinfo {author} {\bibfnamefont {Hao}\ \bibnamefont
  {Li}}, \bibinfo {author} {\bibfnamefont {Yi-Dong}\ \bibnamefont {Huang}},
  \emph {et~al.},\ }\bibfield  {title} {\enquote {\bibinfo {title}
  {Entanglement swapping over 100 km optical fiber with independent entangled
  photon-pair sources},}\ }\href@noop {} {\bibfield  {journal} {\bibinfo
  {journal} {Optica}\ }\textbf {\bibinfo {volume} {4}},\ \bibinfo {pages}
  {1214--1218} (\bibinfo {year} {2017})}\BibitemShut {NoStop}%
\bibitem [{\citenamefont {Shen}\ \emph {et~al.}(2023)\citenamefont {Shen},
  \citenamefont {Yuan}, \citenamefont {Zhang}, \citenamefont {Yu},
  \citenamefont {Zhang}, \citenamefont {Yang}, \citenamefont {Li},
  \citenamefont {Wang}, \citenamefont {Wang}, \citenamefont {Deng} \emph
  {et~al.}}]{shen2023hertz}%
  \BibitemOpen
  \bibfield  {author} {\bibinfo {author} {\bibfnamefont {Si}~\bibnamefont
  {Shen}}, \bibinfo {author} {\bibfnamefont {Chenzhi}\ \bibnamefont {Yuan}},
  \bibinfo {author} {\bibfnamefont {Zichang}\ \bibnamefont {Zhang}}, \bibinfo
  {author} {\bibfnamefont {Hao}\ \bibnamefont {Yu}}, \bibinfo {author}
  {\bibfnamefont {Ruiming}\ \bibnamefont {Zhang}}, \bibinfo {author}
  {\bibfnamefont {Chuanrong}\ \bibnamefont {Yang}}, \bibinfo {author}
  {\bibfnamefont {Hao}\ \bibnamefont {Li}}, \bibinfo {author} {\bibfnamefont
  {Zhen}\ \bibnamefont {Wang}}, \bibinfo {author} {\bibfnamefont {You}\
  \bibnamefont {Wang}}, \bibinfo {author} {\bibfnamefont {Guangwei}\
  \bibnamefont {Deng}},  \emph {et~al.},\ }\bibfield  {title} {\enquote
  {\bibinfo {title} {Hertz-rate metropolitan quantum teleportation},}\
  }\href@noop {} {\bibfield  {journal} {\bibinfo  {journal} {Light: Science \&
  Applications}\ }\textbf {\bibinfo {volume} {12}},\ \bibinfo {pages} {115}
  (\bibinfo {year} {2023})}\BibitemShut {NoStop}%
\bibitem [{\citenamefont {Knaut}\ \emph {et~al.}(2024)\citenamefont {Knaut},
  \citenamefont {Suleymanzade}, \citenamefont {Wei}, \citenamefont {Assumpcao},
  \citenamefont {Stas}, \citenamefont {Huan}, \citenamefont {Machielse},
  \citenamefont {Knall}, \citenamefont {Sutula}, \citenamefont {Baranes} \emph
  {et~al.}}]{knaut2024entanglement}%
  \BibitemOpen
  \bibfield  {author} {\bibinfo {author} {\bibfnamefont {CM}~\bibnamefont
  {Knaut}}, \bibinfo {author} {\bibfnamefont {A}~\bibnamefont {Suleymanzade}},
  \bibinfo {author} {\bibfnamefont {Y-C}\ \bibnamefont {Wei}}, \bibinfo
  {author} {\bibfnamefont {DR}~\bibnamefont {Assumpcao}}, \bibinfo {author}
  {\bibfnamefont {P-J}\ \bibnamefont {Stas}}, \bibinfo {author} {\bibfnamefont
  {YQ}~\bibnamefont {Huan}}, \bibinfo {author} {\bibfnamefont {B}~\bibnamefont
  {Machielse}}, \bibinfo {author} {\bibfnamefont {EN}~\bibnamefont {Knall}},
  \bibinfo {author} {\bibfnamefont {M}~\bibnamefont {Sutula}}, \bibinfo
  {author} {\bibfnamefont {G}~\bibnamefont {Baranes}},  \emph {et~al.},\
  }\bibfield  {title} {\enquote {\bibinfo {title} {Entanglement of nanophotonic
  quantum memory nodes in a telecom network},}\ }\href@noop {} {\bibfield
  {journal} {\bibinfo  {journal} {Nature}\ }\textbf {\bibinfo {volume} {629}},\
  \bibinfo {pages} {573--578} (\bibinfo {year} {2024})}\BibitemShut {NoStop}%
\bibitem [{\citenamefont {L{\"u}tkenhaus}\ \emph {et~al.}(1999)\citenamefont
  {L{\"u}tkenhaus}, \citenamefont {Calsamiglia},\ and\ \citenamefont
  {Suominen}}]{lutkenhaus1999bell}%
  \BibitemOpen
  \bibfield  {author} {\bibinfo {author} {\bibfnamefont {Norbert}\ \bibnamefont
  {L{\"u}tkenhaus}}, \bibinfo {author} {\bibfnamefont {John}\ \bibnamefont
  {Calsamiglia}}, \ and\ \bibinfo {author} {\bibfnamefont {K-A}\ \bibnamefont
  {Suominen}},\ }\bibfield  {title} {\enquote {\bibinfo {title} {Bell
  measurements for teleportation},}\ }\href@noop {} {\bibfield  {journal}
  {\bibinfo  {journal} {Physical Review A}\ }\textbf {\bibinfo {volume} {59}},\
  \bibinfo {pages} {3295} (\bibinfo {year} {1999})}\BibitemShut {NoStop}%
\bibitem [{\citenamefont {Kim}\ \emph {et~al.}(2001)\citenamefont {Kim},
  \citenamefont {Kulik},\ and\ \citenamefont {Shih}}]{kim2001quantum}%
  \BibitemOpen
  \bibfield  {author} {\bibinfo {author} {\bibfnamefont {Yoon-Ho}\ \bibnamefont
  {Kim}}, \bibinfo {author} {\bibfnamefont {Sergei~P}\ \bibnamefont {Kulik}}, \
  and\ \bibinfo {author} {\bibfnamefont {Yanhua}\ \bibnamefont {Shih}},\
  }\bibfield  {title} {\enquote {\bibinfo {title} {Quantum teleportation of a
  polarization state with a complete bell state measurement},}\ }\href@noop {}
  {\bibfield  {journal} {\bibinfo  {journal} {Physical Review Letters}\
  }\textbf {\bibinfo {volume} {86}},\ \bibinfo {pages} {1370} (\bibinfo {year}
  {2001})}\BibitemShut {NoStop}%
\bibitem [{\citenamefont {Sangouard}\ \emph
  {et~al.}(2011{\natexlab{b}})\citenamefont {Sangouard}, \citenamefont
  {Sanguinetti}, \citenamefont {Curtz}, \citenamefont {Gisin}, \citenamefont
  {Thew},\ and\ \citenamefont {Zbinden}}]{sangouard2011faithful}%
  \BibitemOpen
  \bibfield  {author} {\bibinfo {author} {\bibfnamefont {Nicolas}\ \bibnamefont
  {Sangouard}}, \bibinfo {author} {\bibfnamefont {Bruno}\ \bibnamefont
  {Sanguinetti}}, \bibinfo {author} {\bibfnamefont {No{\'e}}\ \bibnamefont
  {Curtz}}, \bibinfo {author} {\bibfnamefont {Nicolas}\ \bibnamefont {Gisin}},
  \bibinfo {author} {\bibfnamefont {Rob}\ \bibnamefont {Thew}}, \ and\ \bibinfo
  {author} {\bibfnamefont {Hugo}\ \bibnamefont {Zbinden}},\ }\bibfield  {title}
  {\enquote {\bibinfo {title} {Faithful entanglement swapping based on
  sum-frequency generation},}\ }\href@noop {} {\bibfield  {journal} {\bibinfo
  {journal} {Physical Review Letters}\ }\textbf {\bibinfo {volume} {106}},\
  \bibinfo {pages} {120403} (\bibinfo {year} {2011}{\natexlab{b}})}\BibitemShut
  {NoStop}%
\bibitem [{\citenamefont {Sephton}\ \emph {et~al.}(2023)\citenamefont
  {Sephton}, \citenamefont {Vall{\'e}s}, \citenamefont {Nape}, \citenamefont
  {Cox}, \citenamefont {Steinlechner}, \citenamefont {Konrad}, \citenamefont
  {Torres}, \citenamefont {Roux},\ and\ \citenamefont
  {Forbes}}]{sephton2023quantum}%
  \BibitemOpen
  \bibfield  {author} {\bibinfo {author} {\bibfnamefont {Bereneice}\
  \bibnamefont {Sephton}}, \bibinfo {author} {\bibfnamefont {Adam}\
  \bibnamefont {Vall{\'e}s}}, \bibinfo {author} {\bibfnamefont {Isaac}\
  \bibnamefont {Nape}}, \bibinfo {author} {\bibfnamefont {Mitchell~A}\
  \bibnamefont {Cox}}, \bibinfo {author} {\bibfnamefont {Fabian}\ \bibnamefont
  {Steinlechner}}, \bibinfo {author} {\bibfnamefont {Thomas}\ \bibnamefont
  {Konrad}}, \bibinfo {author} {\bibfnamefont {Juan~P}\ \bibnamefont {Torres}},
  \bibinfo {author} {\bibfnamefont {Filippus~S}\ \bibnamefont {Roux}}, \ and\
  \bibinfo {author} {\bibfnamefont {Andrew}\ \bibnamefont {Forbes}},\
  }\bibfield  {title} {\enquote {\bibinfo {title} {Quantum transport of
  high-dimensional spatial information with a nonlinear detector},}\
  }\href@noop {} {\bibfield  {journal} {\bibinfo  {journal} {Nature
  Communications}\ }\textbf {\bibinfo {volume} {14}},\ \bibinfo {pages} {8243}
  (\bibinfo {year} {2023})}\BibitemShut {NoStop}%
\bibitem [{\citenamefont {Qiu}\ \emph {et~al.}(2023)\citenamefont {Qiu},
  \citenamefont {Guo},\ and\ \citenamefont {Chen}}]{qiu2023remote}%
  \BibitemOpen
  \bibfield  {author} {\bibinfo {author} {\bibfnamefont {Xiaodong}\
  \bibnamefont {Qiu}}, \bibinfo {author} {\bibfnamefont {Haoxu}\ \bibnamefont
  {Guo}}, \ and\ \bibinfo {author} {\bibfnamefont {Lixiang}\ \bibnamefont
  {Chen}},\ }\bibfield  {title} {\enquote {\bibinfo {title} {Remote transport
  of high-dimensional orbital angular momentum states and ghost images via
  spatial-mode-engineered frequency conversion},}\ }\href@noop {} {\bibfield
  {journal} {\bibinfo  {journal} {Nature Communications}\ }\textbf {\bibinfo
  {volume} {14}},\ \bibinfo {pages} {8244} (\bibinfo {year}
  {2023})}\BibitemShut {NoStop}%
\bibitem [{\citenamefont {Tanzilli}\ \emph {et~al.}(2005)\citenamefont
  {Tanzilli}, \citenamefont {Tittel}, \citenamefont {Halder}, \citenamefont
  {Alibart}, \citenamefont {Baldi}, \citenamefont {Gisin},\ and\ \citenamefont
  {Zbinden}}]{tanzilli2005photonic}%
  \BibitemOpen
  \bibfield  {author} {\bibinfo {author} {\bibfnamefont {Sebastien}\
  \bibnamefont {Tanzilli}}, \bibinfo {author} {\bibfnamefont {Wolfgang}\
  \bibnamefont {Tittel}}, \bibinfo {author} {\bibfnamefont {Matthaeus}\
  \bibnamefont {Halder}}, \bibinfo {author} {\bibfnamefont {Olivier}\
  \bibnamefont {Alibart}}, \bibinfo {author} {\bibfnamefont {Pascal}\
  \bibnamefont {Baldi}}, \bibinfo {author} {\bibfnamefont {Nicolas}\
  \bibnamefont {Gisin}}, \ and\ \bibinfo {author} {\bibfnamefont {Hugo}\
  \bibnamefont {Zbinden}},\ }\bibfield  {title} {\enquote {\bibinfo {title} {A
  photonic quantum information interface},}\ }\href@noop {} {\bibfield
  {journal} {\bibinfo  {journal} {Nature}\ }\textbf {\bibinfo {volume} {437}},\
  \bibinfo {pages} {116--120} (\bibinfo {year} {2005})}\BibitemShut {NoStop}%
\bibitem [{\citenamefont {Fisher}\ \emph {et~al.}(2021)\citenamefont {Fisher},
  \citenamefont {Cernansky}, \citenamefont {Haylock},\ and\ \citenamefont
  {Lobino}}]{fisher2021single}%
  \BibitemOpen
  \bibfield  {author} {\bibinfo {author} {\bibfnamefont {Paul}\ \bibnamefont
  {Fisher}}, \bibinfo {author} {\bibfnamefont {Robert}\ \bibnamefont
  {Cernansky}}, \bibinfo {author} {\bibfnamefont {Ben}\ \bibnamefont
  {Haylock}}, \ and\ \bibinfo {author} {\bibfnamefont {Mirko}\ \bibnamefont
  {Lobino}},\ }\bibfield  {title} {\enquote {\bibinfo {title} {Single photon
  frequency conversion for frequency multiplexed quantum networks in the
  telecom band},}\ }\href@noop {} {\bibfield  {journal} {\bibinfo  {journal}
  {Physical Review Letters}\ }\textbf {\bibinfo {volume} {127}},\ \bibinfo
  {pages} {023602} (\bibinfo {year} {2021})}\BibitemShut {NoStop}%
\bibitem [{\citenamefont {Guerreiro}\ \emph {et~al.}(2013)\citenamefont
  {Guerreiro}, \citenamefont {Pomarico}, \citenamefont {Sanguinetti},
  \citenamefont {Sangouard}, \citenamefont {Pelc}, \citenamefont {Langrock},
  \citenamefont {Fejer}, \citenamefont {Zbinden}, \citenamefont {Thew},\ and\
  \citenamefont {Gisin}}]{guerreiro2013interaction}%
  \BibitemOpen
  \bibfield  {author} {\bibinfo {author} {\bibfnamefont {Thiago}\ \bibnamefont
  {Guerreiro}}, \bibinfo {author} {\bibfnamefont {Enrico}\ \bibnamefont
  {Pomarico}}, \bibinfo {author} {\bibfnamefont {Bruno}\ \bibnamefont
  {Sanguinetti}}, \bibinfo {author} {\bibfnamefont {Nicolas}\ \bibnamefont
  {Sangouard}}, \bibinfo {author} {\bibfnamefont {JS}~\bibnamefont {Pelc}},
  \bibinfo {author} {\bibfnamefont {C}~\bibnamefont {Langrock}}, \bibinfo
  {author} {\bibfnamefont {MM}~\bibnamefont {Fejer}}, \bibinfo {author}
  {\bibfnamefont {Hugo}\ \bibnamefont {Zbinden}}, \bibinfo {author}
  {\bibfnamefont {Robert~T}\ \bibnamefont {Thew}}, \ and\ \bibinfo {author}
  {\bibfnamefont {Nicolas}\ \bibnamefont {Gisin}},\ }\bibfield  {title}
  {\enquote {\bibinfo {title} {Interaction of independent single photons based
  on integrated nonlinear optics},}\ }\href@noop {} {\bibfield  {journal}
  {\bibinfo  {journal} {Nature Communications}\ }\textbf {\bibinfo {volume}
  {4}},\ \bibinfo {pages} {2324} (\bibinfo {year} {2013})}\BibitemShut
  {NoStop}%
\bibitem [{\citenamefont {Guerreiro}\ \emph {et~al.}(2014)\citenamefont
  {Guerreiro}, \citenamefont {Martin}, \citenamefont {Sanguinetti},
  \citenamefont {Pelc}, \citenamefont {Langrock}, \citenamefont {Fejer},
  \citenamefont {Gisin}, \citenamefont {Zbinden}, \citenamefont {Sangouard},\
  and\ \citenamefont {Thew}}]{guerreiro2014nonlinear}%
  \BibitemOpen
  \bibfield  {author} {\bibinfo {author} {\bibfnamefont {Thiago}\ \bibnamefont
  {Guerreiro}}, \bibinfo {author} {\bibfnamefont {A}~\bibnamefont {Martin}},
  \bibinfo {author} {\bibfnamefont {B}~\bibnamefont {Sanguinetti}}, \bibinfo
  {author} {\bibfnamefont {JS}~\bibnamefont {Pelc}}, \bibinfo {author}
  {\bibfnamefont {C}~\bibnamefont {Langrock}}, \bibinfo {author} {\bibfnamefont
  {MM}~\bibnamefont {Fejer}}, \bibinfo {author} {\bibfnamefont {N}~\bibnamefont
  {Gisin}}, \bibinfo {author} {\bibfnamefont {H}~\bibnamefont {Zbinden}},
  \bibinfo {author} {\bibfnamefont {N}~\bibnamefont {Sangouard}}, \ and\
  \bibinfo {author} {\bibfnamefont {RT}~\bibnamefont {Thew}},\ }\bibfield
  {title} {\enquote {\bibinfo {title} {Nonlinear interaction between single
  photons},}\ }\href@noop {} {\bibfield  {journal} {\bibinfo  {journal}
  {Physical Review Letters}\ }\textbf {\bibinfo {volume} {113}},\ \bibinfo
  {pages} {173601} (\bibinfo {year} {2014})}\BibitemShut {NoStop}%
\bibitem [{\citenamefont {Tsujimoto}\ \emph {et~al.}(2024)\citenamefont
  {Tsujimoto}, \citenamefont {Wakui}, \citenamefont {Kishimoto}, \citenamefont
  {Miki}, \citenamefont {Yabuno}, \citenamefont {Terai}, \citenamefont
  {Fujiwara},\ and\ \citenamefont {Kato}}]{tsujimoto2024experimental}%
  \BibitemOpen
  \bibfield  {author} {\bibinfo {author} {\bibfnamefont {Yoshiaki}\
  \bibnamefont {Tsujimoto}}, \bibinfo {author} {\bibfnamefont {Kentaro}\
  \bibnamefont {Wakui}}, \bibinfo {author} {\bibfnamefont {Tadashi}\
  \bibnamefont {Kishimoto}}, \bibinfo {author} {\bibfnamefont {Shigehito}\
  \bibnamefont {Miki}}, \bibinfo {author} {\bibfnamefont {Masahiro}\
  \bibnamefont {Yabuno}}, \bibinfo {author} {\bibfnamefont {Hirotaka}\
  \bibnamefont {Terai}}, \bibinfo {author} {\bibfnamefont {Mikio}\ \bibnamefont
  {Fujiwara}}, \ and\ \bibinfo {author} {\bibfnamefont {Go}~\bibnamefont
  {Kato}},\ }\bibfield  {title} {\enquote {\bibinfo {title} {Experimental
  entanglement swapping through single-photon $\chi^{(2)}$ nonlinearity},}\
  }\href@noop {} {\bibfield  {journal} {\bibinfo  {journal} {arXiv preprint
  arXiv:2411.17267}\ } (\bibinfo {year} {2024})}\BibitemShut {NoStop}%
\bibitem [{\citenamefont {Lu}\ \emph {et~al.}(2020)\citenamefont {Lu},
  \citenamefont {Li}, \citenamefont {Zou}, \citenamefont {Al~Sayem},\ and\
  \citenamefont {Tang}}]{lu2020toward}%
  \BibitemOpen
  \bibfield  {author} {\bibinfo {author} {\bibfnamefont {Juanjuan}\
  \bibnamefont {Lu}}, \bibinfo {author} {\bibfnamefont {Ming}\ \bibnamefont
  {Li}}, \bibinfo {author} {\bibfnamefont {Chang-Ling}\ \bibnamefont {Zou}},
  \bibinfo {author} {\bibfnamefont {Ayed}\ \bibnamefont {Al~Sayem}}, \ and\
  \bibinfo {author} {\bibfnamefont {Hong~X}\ \bibnamefont {Tang}},\ }\bibfield
  {title} {\enquote {\bibinfo {title} {Toward 1\% single-photon anharmonicity
  with periodically poled lithium niobate microring resonators},}\ }\href@noop
  {} {\bibfield  {journal} {\bibinfo  {journal} {Optica}\ }\textbf {\bibinfo
  {volume} {7}},\ \bibinfo {pages} {1654--1659} (\bibinfo {year}
  {2020})}\BibitemShut {NoStop}%
\bibitem [{\citenamefont {Zhao}\ and\ \citenamefont
  {Fang}(2022)}]{zhao2022ingap}%
  \BibitemOpen
  \bibfield  {author} {\bibinfo {author} {\bibfnamefont {Mengdi}\ \bibnamefont
  {Zhao}}\ and\ \bibinfo {author} {\bibfnamefont {Kejie}\ \bibnamefont
  {Fang}},\ }\bibfield  {title} {\enquote {\bibinfo {title} {In{G}a{P} quantum
  nanophotonic integrated circuits with 1.5\% nonlinearity-to-loss ratio},}\
  }\href@noop {} {\bibfield  {journal} {\bibinfo  {journal} {Optica}\ }\textbf
  {\bibinfo {volume} {9}},\ \bibinfo {pages} {258--263} (\bibinfo {year}
  {2022})}\BibitemShut {NoStop}%
\bibitem [{\citenamefont {Akin}\ \emph {et~al.}(2025)\citenamefont {Akin},
  \citenamefont {Zhao}, \citenamefont {Kwiat}, \citenamefont {Goldschmidt},\
  and\ \citenamefont {Fang}}]{akin2025faithful}%
  \BibitemOpen
  \bibfield  {author} {\bibinfo {author} {\bibfnamefont {Joshua}\ \bibnamefont
  {Akin}}, \bibinfo {author} {\bibfnamefont {Yunlei}\ \bibnamefont {Zhao}},
  \bibinfo {author} {\bibfnamefont {Paul~G}\ \bibnamefont {Kwiat}}, \bibinfo
  {author} {\bibfnamefont {Elizabeth~A}\ \bibnamefont {Goldschmidt}}, \ and\
  \bibinfo {author} {\bibfnamefont {Kejie}\ \bibnamefont {Fang}},\ }\bibfield
  {title} {\enquote {\bibinfo {title} {Faithful quantum teleportation via a
  nanophotonic nonlinear bell state analyzer},}\ }\href@noop {} {\bibfield
  {journal} {\bibinfo  {journal} {Physical Review Letters}\ }\textbf {\bibinfo
  {volume} {134}},\ \bibinfo {pages} {160802} (\bibinfo {year}
  {2025})}\BibitemShut {NoStop}%
\bibitem [{\citenamefont {{\'S}liwa}\ and\ \citenamefont
  {Banaszek}(2003)}]{sliwa2003conditional}%
  \BibitemOpen
  \bibfield  {author} {\bibinfo {author} {\bibfnamefont {Cezary}\ \bibnamefont
  {{\'S}liwa}}\ and\ \bibinfo {author} {\bibfnamefont {Konrad}\ \bibnamefont
  {Banaszek}},\ }\bibfield  {title} {\enquote {\bibinfo {title} {Conditional
  preparation of maximal polarization entanglement},}\ }\href@noop {}
  {\bibfield  {journal} {\bibinfo  {journal} {Physical Review A}\ }\textbf
  {\bibinfo {volume} {67}},\ \bibinfo {pages} {030101} (\bibinfo {year}
  {2003})}\BibitemShut {NoStop}%
\bibitem [{\citenamefont {Pittman}\ \emph {et~al.}(2003)\citenamefont
  {Pittman}, \citenamefont {Donegan}, \citenamefont {Fitch}, \citenamefont
  {Jacobs}, \citenamefont {Franson}, \citenamefont {Kok}, \citenamefont {Lee},\
  and\ \citenamefont {Dowling}}]{pittman2003heralded}%
  \BibitemOpen
  \bibfield  {author} {\bibinfo {author} {\bibfnamefont {Todd~B}\ \bibnamefont
  {Pittman}}, \bibinfo {author} {\bibfnamefont {Michelle~M}\ \bibnamefont
  {Donegan}}, \bibinfo {author} {\bibfnamefont {Michael~J}\ \bibnamefont
  {Fitch}}, \bibinfo {author} {\bibfnamefont {Bryan~C}\ \bibnamefont {Jacobs}},
  \bibinfo {author} {\bibfnamefont {James~D}\ \bibnamefont {Franson}}, \bibinfo
  {author} {\bibfnamefont {Pieter}\ \bibnamefont {Kok}}, \bibinfo {author}
  {\bibfnamefont {Hwang}\ \bibnamefont {Lee}}, \ and\ \bibinfo {author}
  {\bibfnamefont {Jonathan~P}\ \bibnamefont {Dowling}},\ }\bibfield  {title}
  {\enquote {\bibinfo {title} {Heralded two-photon entanglement from
  probabilistic quantum logic operations on multiple parametric down-conversion
  sources},}\ }\href@noop {} {\bibfield  {journal} {\bibinfo  {journal} {IEEE
  Journal of Selected Topics in Quantum Electronics}\ }\textbf {\bibinfo
  {volume} {9}},\ \bibinfo {pages} {1478--1482} (\bibinfo {year}
  {2003})}\BibitemShut {NoStop}%
\bibitem [{\citenamefont {Barz}\ \emph {et~al.}(2010)\citenamefont {Barz},
  \citenamefont {Cronenberg}, \citenamefont {Zeilinger},\ and\ \citenamefont
  {Walther}}]{barz2010heralded}%
  \BibitemOpen
  \bibfield  {author} {\bibinfo {author} {\bibfnamefont {Stefanie}\
  \bibnamefont {Barz}}, \bibinfo {author} {\bibfnamefont {Gunther}\
  \bibnamefont {Cronenberg}}, \bibinfo {author} {\bibfnamefont {Anton}\
  \bibnamefont {Zeilinger}}, \ and\ \bibinfo {author} {\bibfnamefont {Philip}\
  \bibnamefont {Walther}},\ }\bibfield  {title} {\enquote {\bibinfo {title}
  {Heralded generation of entangled photon pairs},}\ }\href@noop {} {\bibfield
  {journal} {\bibinfo  {journal} {Nature Photonics}\ }\textbf {\bibinfo
  {volume} {4}},\ \bibinfo {pages} {553--556} (\bibinfo {year}
  {2010})}\BibitemShut {NoStop}%
\bibitem [{\citenamefont {Wagenknecht}\ \emph {et~al.}(2010)\citenamefont
  {Wagenknecht}, \citenamefont {Li}, \citenamefont {Reingruber}, \citenamefont
  {Bao}, \citenamefont {Goebel}, \citenamefont {Chen}, \citenamefont {Zhang},
  \citenamefont {Chen},\ and\ \citenamefont
  {Pan}}]{wagenknecht2010experimental}%
  \BibitemOpen
  \bibfield  {author} {\bibinfo {author} {\bibfnamefont {Claudia}\ \bibnamefont
  {Wagenknecht}}, \bibinfo {author} {\bibfnamefont {Che-Ming}\ \bibnamefont
  {Li}}, \bibinfo {author} {\bibfnamefont {Andreas}\ \bibnamefont
  {Reingruber}}, \bibinfo {author} {\bibfnamefont {Xiao-Hui}\ \bibnamefont
  {Bao}}, \bibinfo {author} {\bibfnamefont {Alexander}\ \bibnamefont {Goebel}},
  \bibinfo {author} {\bibfnamefont {Yu-Ao}\ \bibnamefont {Chen}}, \bibinfo
  {author} {\bibfnamefont {Qiang}\ \bibnamefont {Zhang}}, \bibinfo {author}
  {\bibfnamefont {Kai}\ \bibnamefont {Chen}}, \ and\ \bibinfo {author}
  {\bibfnamefont {Jian-Wei}\ \bibnamefont {Pan}},\ }\bibfield  {title}
  {\enquote {\bibinfo {title} {Experimental demonstration of a heralded
  entanglement source},}\ }\href@noop {} {\bibfield  {journal} {\bibinfo
  {journal} {Nature Photonics}\ }\textbf {\bibinfo {volume} {4}},\ \bibinfo
  {pages} {549--552} (\bibinfo {year} {2010})}\BibitemShut {NoStop}%
\bibitem [{\citenamefont {Braunstein}\ and\ \citenamefont
  {Van~Loock}(2005)}]{braunstein2005quantum}%
  \BibitemOpen
  \bibfield  {author} {\bibinfo {author} {\bibfnamefont {Samuel~L}\
  \bibnamefont {Braunstein}}\ and\ \bibinfo {author} {\bibfnamefont {Peter}\
  \bibnamefont {Van~Loock}},\ }\bibfield  {title} {\enquote {\bibinfo {title}
  {Quantum information with continuous variables},}\ }\href@noop {} {\bibfield
  {journal} {\bibinfo  {journal} {Reviews of Modern Physics}\ }\textbf
  {\bibinfo {volume} {77}},\ \bibinfo {pages} {513--577} (\bibinfo {year}
  {2005})}\BibitemShut {NoStop}%
\bibitem [{\citenamefont {Yuan}\ \emph {et~al.}(2010)\citenamefont {Yuan},
  \citenamefont {Bao}, \citenamefont {Lu}, \citenamefont {Zhang}, \citenamefont
  {Peng},\ and\ \citenamefont {Pan}}]{yuan2010entangled}%
  \BibitemOpen
  \bibfield  {author} {\bibinfo {author} {\bibfnamefont {Zhen-Sheng}\
  \bibnamefont {Yuan}}, \bibinfo {author} {\bibfnamefont {Xiao-Hui}\
  \bibnamefont {Bao}}, \bibinfo {author} {\bibfnamefont {Chao-Yang}\
  \bibnamefont {Lu}}, \bibinfo {author} {\bibfnamefont {Jun}\ \bibnamefont
  {Zhang}}, \bibinfo {author} {\bibfnamefont {Cheng-Zhi}\ \bibnamefont {Peng}},
  \ and\ \bibinfo {author} {\bibfnamefont {Jian-Wei}\ \bibnamefont {Pan}},\
  }\bibfield  {title} {\enquote {\bibinfo {title} {Entangled photons and
  quantum communication},}\ }\href@noop {} {\bibfield  {journal} {\bibinfo
  {journal} {Physics Reports}\ }\textbf {\bibinfo {volume} {497}},\ \bibinfo
  {pages} {1--40} (\bibinfo {year} {2010})}\BibitemShut {NoStop}%
\bibitem [{\citenamefont {Kim}\ \emph {et~al.}(2006)\citenamefont {Kim},
  \citenamefont {Fiorentino},\ and\ \citenamefont {Wong}}]{kim2006phase}%
  \BibitemOpen
  \bibfield  {author} {\bibinfo {author} {\bibfnamefont {Taehyun}\ \bibnamefont
  {Kim}}, \bibinfo {author} {\bibfnamefont {Marco}\ \bibnamefont {Fiorentino}},
  \ and\ \bibinfo {author} {\bibfnamefont {Franco~NC}\ \bibnamefont {Wong}},\
  }\bibfield  {title} {\enquote {\bibinfo {title} {Phase-stable source of
  polarization-entangled photons using a polarization {S}agnac
  interferometer},}\ }\href@noop {} {\bibfield  {journal} {\bibinfo  {journal}
  {Physical Review A}\ }\textbf {\bibinfo {volume} {73}},\ \bibinfo {pages}
  {012316} (\bibinfo {year} {2006})}\BibitemShut {NoStop}%
\bibitem [{\citenamefont {Akin}\ \emph {et~al.}(2024)\citenamefont {Akin},
  \citenamefont {Zhao}, \citenamefont {Misra}, \citenamefont {Haque},\ and\
  \citenamefont {Fang}}]{akin2024ingap}%
  \BibitemOpen
  \bibfield  {author} {\bibinfo {author} {\bibfnamefont {Joshua}\ \bibnamefont
  {Akin}}, \bibinfo {author} {\bibfnamefont {Yunlei}\ \bibnamefont {Zhao}},
  \bibinfo {author} {\bibfnamefont {Yuvraj}\ \bibnamefont {Misra}}, \bibinfo
  {author} {\bibfnamefont {AKM~Naziul}\ \bibnamefont {Haque}}, \ and\ \bibinfo
  {author} {\bibfnamefont {Kejie}\ \bibnamefont {Fang}},\ }\bibfield  {title}
  {\enquote {\bibinfo {title} {In{G}a{P} $\chi^{(2)}$ integrated photonics
  platform for broadband, ultra-efficient nonlinear conversion and entangled
  photon generation},}\ }\href@noop {} {\bibfield  {journal} {\bibinfo
  {journal} {Light: Science \& Applications}\ }\textbf {\bibinfo {volume}
  {13}},\ \bibinfo {pages} {290} (\bibinfo {year} {2024})}\BibitemShut
  {NoStop}%
\bibitem [{\citenamefont {Boyd}(2020)}]{boyd2020nonlinear}%
  \BibitemOpen
  \bibfield  {author} {\bibinfo {author} {\bibfnamefont {Robert~W}\
  \bibnamefont {Boyd}},\ }\href@noop {} {\emph {\bibinfo {title} {Nonlinear
  Optics}}}\ (\bibinfo  {publisher} {Academic Press},\ \bibinfo {year}
  {2020})\BibitemShut {NoStop}%
\bibitem [{\citenamefont {Ren}\ \emph {et~al.}(2017)\citenamefont {Ren},
  \citenamefont {Xu}, \citenamefont {Yong}, \citenamefont {Zhang},
  \citenamefont {Liao}, \citenamefont {Yin}, \citenamefont {Liu}, \citenamefont
  {Cai}, \citenamefont {Yang}, \citenamefont {Li} \emph
  {et~al.}}]{ren2017ground}%
  \BibitemOpen
  \bibfield  {author} {\bibinfo {author} {\bibfnamefont {Ji-Gang}\ \bibnamefont
  {Ren}}, \bibinfo {author} {\bibfnamefont {Ping}\ \bibnamefont {Xu}}, \bibinfo
  {author} {\bibfnamefont {Hai-Lin}\ \bibnamefont {Yong}}, \bibinfo {author}
  {\bibfnamefont {Liang}\ \bibnamefont {Zhang}}, \bibinfo {author}
  {\bibfnamefont {Sheng-Kai}\ \bibnamefont {Liao}}, \bibinfo {author}
  {\bibfnamefont {Juan}\ \bibnamefont {Yin}}, \bibinfo {author} {\bibfnamefont
  {Wei-Yue}\ \bibnamefont {Liu}}, \bibinfo {author} {\bibfnamefont {Wen-Qi}\
  \bibnamefont {Cai}}, \bibinfo {author} {\bibfnamefont {Meng}\ \bibnamefont
  {Yang}}, \bibinfo {author} {\bibfnamefont {Li}~\bibnamefont {Li}},  \emph
  {et~al.},\ }\bibfield  {title} {\enquote {\bibinfo {title}
  {Ground-to-satellite quantum teleportation},}\ }\href@noop {} {\bibfield
  {journal} {\bibinfo  {journal} {Nature}\ }\textbf {\bibinfo {volume} {549}},\
  \bibinfo {pages} {70--73} (\bibinfo {year} {2017})}\BibitemShut {NoStop}%
\end{thebibliography}
\end{document}